\documentclass[twocolumn,showpacs,twoside,pra]{revtex4}
\usepackage{amssymb, amsbsy, amsmath, latexsym, dsfont, array, layout, graphicx}
\usepackage{color}

\newcommand{\ket}[1]{\left|{#1}\right\rangle}
\newcommand{\bra}[1]{\left\langle{#1}\right|}

\begin{document}

\title{Implementation of multi-walker quantum walks with cavity grid}
\author{Peng Xue}
\affiliation{Department of Physics, Southeast University, Nanjing
211189 China}
\date{\today}

\begin{abstract}
We show how multi-walker quantum walks can be implemented in a
quantum quincunx created via cavity quantum electrodynamics. The
implementation of a quantum walk with a multi-walker opens up the
interesting possibility to introduce entanglement and more advanced
walks. With different coin tosses and initial states the
multi-walker quantum walk shows different probability distributions
which deviate strongly from the classical random walks with
quadratic enhanced spreadings and localization effects. By
introducing decoherence, the transition from quantum walks to the
classical versions is observed. We introduce the average fidelity
decay as a signature to investigate the decoherence-induced
irreversibility of quantum walks.
\end{abstract}


\maketitle

\section{Introduction}

Quantum walks (QWs)~\cite{QW} offer an alternative approach to
implement quantum algorithms~\cite{algorithm} compared to the
typical circuit-based~\cite{circuit} or measurement
based~\cite{oneway} models for quantum algorithms. There are two
types of QWs: the continuous time QWs~\cite{con} and the discrete
time QWs~\cite{dis}. The realization for QWs have been proposed in
quantum optics~\cite{qo}, ion trap~\cite{ion1,ion2}, cavity quantum
electrodynamics (QED)~\cite{cavity1,cavity2} and optical
lattice~\cite{ol} systems in a decade. QWs on single
photons~\cite{expsp}, trapped ions~\cite{expion} and neutral
atoms~\cite{expatom} have been realized in the laboratory,
respectively. More recently, higher dimensional QWs \cite{Mackay02}
and QWs involving more particles
\cite{more1,more2,more3,more4,more5,more6,more7,more8} are studied.
This reveals the additional features offered by quantum mechanics,
such as quantum correlations~\cite{more8} and indistinguishability.
Compared to the proposals on QWs with single-walker, we extend the
QWs by using multi-walker. The implementation of a QW with a
multi-walker opens up the interesting possibility to introduce
entanglement and more advanced walks.

To create a QW, the following steps are followed: a particular QW is
chosen, in our case a discrete multi-walker QW with joint coins each
of which decides to the corresponding walker's position shifts.
Furthermore, a signature for QW behavior is identified, such as
enhanced diffusion or uniformity of the distribution for the
walkers' degree of freedom. Then a physical system is chosen whose
Hamiltonian dynamics match the evolution of the QW. Finally, open
system dynamics are incorporated into the analysis in order to
account for non-unitary evolution as well as to incorporate
realistic measurement into the model.

Compared to random walks (RWs), QWs are reversible. The
irreversibility due to decoherence transmits the QW to RW . It is of
great interest to show the variation of the irreversibility in the
time evolution of the QW. The probability distribution and standard
deviation of the distribution are used to study the irreversibility
of QWs in the present of decoherence. However those approaches are
not operational. That means neither of them provides a way for
direct experimental observation of the irreversibility of the QW. In
the present of decoherence, except for the probability distribution
and the standard deviation, we introduce an operational
measurement---the average fidelity decay (AFD)~\cite{Feng,AFD}. The
AFD can reveal the response of the system to the decoherence which
is closely related to the properties of both the system and the
environment, and provide an experimentally available way to monitor
the detrimental influence on the QW with different decoherence
sources.

\section{quantum walks with one- and multi-walker}

Let us first briefly review a single-walker QW. The Hilbert space of
the walker+coin is given by a tensor product
\begin{equation}
\mathcal{H}=\mathcal{H}_\text{w}\otimes\mathcal{H}_\text{c}
\end{equation}
of the walker space $\mathcal{H}_\text{w}$ and the two-dimensional
(2D) coin space
\begin{equation}
\mathcal{H}_\text{c}=\text{span}\{\ket{-1},\ket{1}\}.
\end{equation}
We consider a walker starting the QW from the origin, i.e., the
initial state has the form
\begin{equation}
\ket{\psi}_\text{ini}=\ket{\psi_0}_\text{w}\otimes\ket{\psi}_\text{c},
\end{equation}
where $\ket{\psi_0}_\text{w}$ and $\ket{\psi}_\text{c}$ denote the
initial state of the walker and coin respectively. After $N$ steps
of the QW, the state of the walker+coin is given by
\begin{align}
\ket{\psi_N}&=U^N\ket{\psi}_\text{ini}\\
&=\sum_jp_{-1}(j,N)\ket{\psi_j}_\text{w}\ket{-1}
+p_1(j,N)\ket{\psi_j}_\text{w}\ket{1},\nonumber
\end{align}
where the unitary propagator $U$ has the form
\begin{equation}
U=S(\mathbb{I}\otimes C).
\end{equation}
The probability distribution generated by the QW is given by
\begin{align}
p(j,N)&=\Big|_\text{w}\bra{\psi_j}\bra{-1}\psi_N\rangle\Big|^2+\Big|_\text{w}\bra{\psi_j}\bra{1}\psi_N\rangle\Big|^2\nonumber\\
&=\Big|p_{-1}(j,N)\Big|^2+\Big|p_{1}(j,N)\Big|^2.
\end{align}
The coin operator $C$ flips the state of the coin before the walker
is displaced. In principle, $C$ can be an arbitrary unitary
operation on the coin space $\mathcal{H}_\text{c}$. We choose the
most studied case of the Hadamard coin, denoted by
$H=\begin{pmatrix}
                                                          1 & 1 \\
                                                          1 & -1 \\
                                                        \end{pmatrix}/\sqrt{2}
$, which is defined by its action on the basis states,
\begin{equation}
H\ket{\pm 1}=\big(\ket{-1}\mp\ket{1}\big)/\sqrt{2}.
\end{equation}
After the coin flip, the step operator $S$ displaces the walker from
its current state according to its coin state
\begin{equation}
S\ket{\psi_j}_\text{w}\ket{\pm 1}\longrightarrow\ket{\psi_j\pm
\delta}_\text{w}\ket{\pm 1},
\end{equation}
where $\delta$ is the step size. The coefficients $p_{\pm1}(j,N)$
represent the probability amplitudes of finding the walker at
$\ket{\psi_j}_\text{w}$ after $N$ steps of the QW with the coin
state $\ket{\pm1}$.

As an extension, we choose the $N$-walker QW over circles in phase
space, which arises naturally for $N$ harmonic oscillators. Points
in phase space correspond to the oscillator position-momentum pair
$(x,p)$, which we henceforth refer to as the phase space `location'.

For the discrete $N$-walker QW on the circles, each of the walker's
location as a point in phase space is replaced by a localized
wavefunction centered at location $(x,p)$, and the random flips are
replaced by joint quantum coins given by qubits, which are flipped
by a unitary operation and then entangled with the oscillators by
free evolution. An example of $N$-walker state is the product state
\begin{equation}
\ket{\psi}_\text{w}=\ket{\phi_1}\otimes\ket{\phi_2}\otimes...\otimes\ket{\phi_N}.
\label{eq:walker}
\end{equation}
An example of the coin state is
$\ket{\psi}_\text{c}=\ket{c_1,c_2,...,c_N}$. The archetypal discrete
time $N$-walker QW consists of two building blocks---a coin operator
$C$ and a step operator $S$. The coin is essentially an ancillary
parameter that is used by the step operator to decide how to
propagate the walker. The simplest example of the coin operation is
to apply a Hadamard transformation to each qubit in the
decomposition of Eq.~(\ref{eq:walker}). This internal transformation
is separable, in the sense that it does not produce entanglement
between the spatial degree of freedom. Other choices for the coin
operations include the entangling coin operation, the discrete
Fourier transform (DFT) and the Grover operation. A common choice of
step operator is
\begin{align}
&S\ket{\psi}_\text{w}\ket{\psi}_\text{c}\\
&=\ket{\phi_1+c_1\delta}\otimes\ket{\phi_2+c_2\delta}\otimes...\otimes\ket{\phi_N+c_N\delta}\ket{c_1,c_2,...,c_N},\nonumber
\end{align}
where $c_i\in\{-1,1\}$ and
\begin{equation}\ket{\phi_j}=\frac{1}{\sqrt{M}}\sum_{n=0}^M\text{e}^{i\phi_jn}\ket{n}\end{equation}
for $j=1,...,N$ is the phase state.

Using two-walker QW as an example, one obvious generalization of
coin tossing operations is to apply a Hadamard transformation $H$ to
each qubit of the coin state. This choice can be viewed as two
independent coin tosses on each qubit of the coin state. The
transformation is
\begin{equation}
C_1=H\otimes H=\frac{1}{2}\begin{pmatrix}
                   1 & 1 & 1 & 1 \\
                   1 & -1 & 1 & -1 \\
                   1 & 1 & -1 & -1 \\
                   1 & -1 & -1 & 1 \\
                \end{pmatrix}.
\end{equation}
This coin operation is separable, in the sense that it does not
produce entanglement between the spatial degrees of freedom.

In principle, any unitary transformations on the coin state can
replace Hadamard transformation and be used as coin tosses. Now we
introduce non-separated coin tosses. One obvious generalization of
coin tosses, which is not separable and does produce entanglement
between the coin qubits, is the root of SWAP gate operation, defined
as follows
\begin{equation}
C_2=\sqrt{i\text{SWAP}}=\begin{pmatrix}
                 1 & 0 & 0 & 0 \\
                 0 & \cos\theta & i\sin\theta & 0 \\
                 0 & i\sin\theta & \cos\theta & 0 \\
                 0 & 0 & 0 & 1 \\
              \end{pmatrix}.
\end{equation}

Another choice of non-separated ncoin tosses---the discrete Fourier
transform (DFT) $D$, defined as follows:
\begin{equation}
\label{eq:DFT} C_3=D=\frac{1}{2}\begin{pmatrix}
                 1 & 1 & 1 & 1 \\
                 1 & i &-1 &-i \\
                 1 &-1 & 1 &-1 \\
                 1 &-i &-1 &i \\
              \end{pmatrix}
\end{equation}
transforms any coin translation eigenstate into an equally weighted
superposition of all the eigenstates and entangles the spatial
degrees of freedom of coin qubits.

There are, of course, an infinite variety of other non-separable
choices for the coin tosses by employing different phase
relationships. Finally we introduce the Grover operator as a
non-separated coin toss defined as follows
\begin{equation}
\label{eq:Grover} C_4=G=\frac{1}{2}\begin{pmatrix}
                 -1 & 1 & 1 & 1 \\
                 1 & -1 & 1 & 1 \\
                 1 & 1 & -1 & 1 \\
                 1 & 1 & 1 & -1 \\
              \end{pmatrix}.
\end{equation}

We now investigate the effect of measurement on a two-walker system.
After $k$th steps of two-walker QW, the state of the walker+coin
system becomes $\ket{\psi_k}=(SC)^k\ket{\psi_\text{ini}}$. Suppose
we perform a measurement and detect the first (second) walker at
position $\phi$. Then the state will be projected into
\begin{equation}
\label{eq:distribution}
P_{1(2)}(\phi)=\bra{\phi}\text{Tr}_{1(2)}\rho_\text{w}\ket{\phi},
\end{equation}
where $\rho_\text{w}=\text{Tr}_\text{c}(\ket{\psi_k}\bra{\psi_k})$.

The dispersion of the distribution (\ref{eq:distribution}) is
especially important. As moments are not particulary useful for
distributions over compact domains, other strategies are needed. For
the phase distribution over the domain $\left[0,2\pi\right]$,
Holevo's version~\cite{Holevo} of the standard deviation
\begin{equation}
\sigma(\phi)=\sqrt{\Big|\int_0^{2\pi}\text{d}\phi
P(\phi)\text{e}^{i\phi}\Big|^{-2}-1}
\end{equation}
is
particulary useful as it reduces to the ordinary standard deviation
for small spreads and is sensible when the dispersion is large over
the domain.

We begin our analysis the results of two-walker QW with different
coin tosses. The initial conditions for the coin state were chosen
to be the separable state composed of all qubits in the states
$\ket{1}$, $\big(\ket{1}+\ket{-1}\big)/\sqrt{2}$, and
$\big(\ket{1}+i\ket{-1}\big)/\sqrt{2}$, which lead to three
different probability distributions.

\begin{figure}[tbp]
   \includegraphics[width=8.5cm]{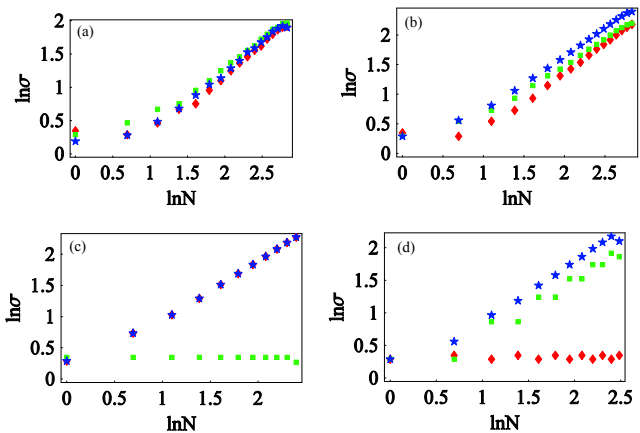}
   \caption{(Color online.) The ln-ln plot of the position spreads $\sigma(\phi)$ of the first walker taking the ideal
   two-walkers QWs with different coin tosses (a) DFT $D$, (b) $H\otimes H$, (c)
   $\sqrt{i\text{SWAP}}$ and (d) Grover $G$ as functions of the
   position $\phi$ for different initial coin states
   $\ket{\Psi}_\text{c1}=\big(\ket{1}+\ket{-1}\big)\otimes\big(\ket{1}+\ket{-1}\big)/2$ ($\blacklozenge$),
   $\ket{\Psi}_\text{c2}=\ket{1}\otimes\ket{1}$ ($\blacksquare$) and
   $\ket{\Psi}_\text{c3}=\big(\ket{1}+i\ket{-1}\big)\otimes\big(\ket{1}+i\ket{-1}\big)/2$
   ($\bigstar$).}
   \label{standard deviations}
\end{figure}

\begin{table}
\begin{tabular}{|c|c|c|c|}
  \hline
  $s\pm\Delta s$ & $\ket{\psi}_\text{c1}$ & $\ket{\psi}_\text{c2}$ & $\ket{\psi}_\text{c3}$ \\\hline
  $D$ & $0.940\pm0.003$ & $0.880\pm0.002$ & $0.887\pm0.003$ \\\hline
  $H\otimes H$ & $0.979\pm0.005$ & $0.897\pm0.007$ & $0.994\pm0.004$ \\\hline
  $\sqrt{i\text{SWAP}}$ & $0.965\pm0.001$ & $0$ & $0.965\pm0.001$ \\\hline
  $G$ & $0$ & $0.911\pm0.002$ & $0.973\pm0.006$ \\\hline
\end{tabular}
\caption{
    Linear regression results for
    $\ln \sigma(\phi)=(s\pm\Delta s)\ln N$ for the ideal case with the step size $\delta=0.8$ in ln-ln scale shown in Fig.~1.}
 \label{table1}
\end{table}

For the case of separable transformation with separable initial
conditions, the different walkers behave independently; thus, the
variance can be expressed in terms of one-walker case. Furthermore,
a bias could be introduced into the transformation and give a
different weighted superposition of translation eigenstates. Thus
the phase distribution of walker also depends on the initial state
of the coin. The time dependence of the standard deviation for QW is
plotted in Fig.~1 and the corresponding slopes $\Delta\sigma/\Delta
t$ ($\ln\sigma(\phi)/\ln N$) are presented in Table~I. We observe
that the standard deviation for QW with the same coin toss depends
on the symmetry of the initial states. However the quantum behavior
can always be observed except for the two cases---the
$\sqrt{\text{iSWAP}}$ QW with the initial coin state
$\ket{1}\otimes\ket{1}$ and the Grover $G$ QW with the initial coin
state
$\left(\ket{1}+\ket{-1}\right)\otimes\left(\ket{1}+\ket{-1}\right)/2$,
which shows the localization effect on QW where the walker's spread
becomes constant \cite{Keating,SC}.

Any real implementation of a quantum system must deal with the issue
of decoherence, which tends to destroy quantum correlations. The
entanglement decay due to noise leads the transition from QWs to
RWs. In the present of decoherence, except for the probability
distribution and the standard deviation of the distribution another
operational measurement here is introduced as a new signature for
QWs. We measure the irreversibility of the QW due to decoherence by
the AFD which is defined based on fidelity decay, that is the square
modulus of the overlap between two time dependent final states with
the same initial state under the time evolution without and with
decoherence respectively. The AFD~\cite{AFD} is defined as
\begin{equation}
AFD=\text{Tr}\big[\rho(t)\ket{\Psi_N}\bra{\Psi_N}\big],
\end{equation}
where $\rho(t)$ is the density matrix of the walker+coin system
after the time evolution. The AFD reveals the response of the
walker+coin system to the decoherence and is sensitive to different
sources of decoherence.

\section{implementation of multi-walker QWs on circles}

Multi-walker QWs can be implemented with scalable cavity grid
\cite{grrid} with superconducting circuits~\cite{Bla04,Bla07}. The
walkers are represented by the cavity modes and the coin states
$\{\ket{-1},\ket{1}\}$ are encoded with charge qubits. The cavity
grid consists of cavity modes belonging to $N_\text{H}$ horizontal
(H) and $N_\text{V}$ vertical (V) cavities,
$\hat{H}_\text{cav}=\sum_{j=1}^{N_\text{H}}\omega_j^H\hat{a}_j^\dagger\hat{a}_j
+\sum_{j=1}^{N_\text{V}}\omega_j^V\hat{b}^\dagger_j\hat{b}_j$,
coupled to one charge qubit at each intersection $(i,j)$,
generalizing to a 2D architecture:
\begin{equation}
\label{eq:cavqb}
\hat{H}_\text{cav-qb}=\sum_{i,j}\ket{1}_{ij}\bra{1}\left[g_{ij}^H(\hat{a}_i+\hat{a}^\dagger_i)+g_{ij}^V(\hat{b}_j+\hat{b}_j^\dagger)\right].
\end{equation}
The coupling $g_{ij}^{H(V)}$ between the horizontal (vertical)
cavity mode $i$ ($j$) and the charge qubits is switchable by the
external electric field on each charge qubit~\cite{Xue}.
Eq.~(\ref{eq:cavqb}) leads to the Jaynes-Cummings (JC) model and the
cavity-mediated interaction between qubits.

For a multi-walker QW, one can implement the coin operation on
charge qubits capacitively coupled to a vertical cavity mode ($j$)
with cavity-assisted interaction. The conditional position shifts of
each walker can be implemented for each horizontal cavity mode $i$
under the free evolution of the JC interactions between horizontal
cavity mode $i$ and charge qubit $(i,j)$.

The multi-walker QW with cavity grid can be implemented by steps as
follows. Step I, the state of the cavity grid is prepared in a
certain initial state. Step II, the couplings between the vertical
cavity mode and the charge qubits are turned on to implement the
coin operation on charge qubits with cavity-assisted interactions.
Step III, the couplings between the vertical cavity mode and the
charge qubits are turned off and those between the horizontal cavity
modes and the charge qubits are turned on to implement the
conditional position shifts of each walker due to the coin states.
Then, we repeat steps II and III for the next step of multi-walker
QW.

\subsection{The Conditional Phase Shifts}

Using a two-walker QW as an example, we consider a system including
two two-level charge qubit coupled to a cavity grid with the
structure mentioned above. The coupling between each charge qubit
and the corresponding horizontal cavity field is used to implement
conditional phase shifts on circles and the charge states are used
to implement quantum coins, each of them with two possible
operations. The physical implementation of the conditional phase
shifts of multi-walker quantum walks can be implemented by the
free-evolution of the cavity-assisted interaction
Eq.~(\ref{eq:cavqb}), which can be rewritten in the JC
model~\cite{Bla04}
\begin{align}
\label{eq:H}
    \hat{H}_\text{JC}
=\sum_{j=1,2}\left[\omega_\text{c}\hat{a}^\dagger_j\hat{a}_j+\frac{\omega_\text{a}}{2}
        \hat{\sigma}_z^j+g(\hat{a}_j^\dagger\hat{\sigma}_-^j+\hat{a}_j\hat{\sigma}_+^j)\right]
\end{align}
with $\omega_\text{a}$ and $\omega_\text{c}$ the coin and cavity
frequencies, respectively, and $g$ the coupling strength. In the
dispersive regime,
\begin{equation}
|\Delta|=|\omega_\text{a}-\omega_\text{c}|\gg g,
\end{equation}
and in a rotating frame, the effective interaction Hamiltonian is
\begin{equation}
\label{eq:Hint}
     \hat{H}_\text{int}=\sum_{j=1,2}\chi\hat{a}^\dagger_j\hat{a}_j\hat{\sigma}_z^j
\end{equation}
with the cavity pull of the resonator
\begin{equation}
\chi=\frac{g^2}{\Delta}. \label{eq:cavity pull}
\end{equation}

After the coin flipping operation, the time evolutions of the
interactive Hamiltonian give the conditional position shifts due to
the charge states
\begin{equation}
U_j=\exp(i\Delta\theta_j\hat{a}^\dagger_j\hat{a}_j\hat{\sigma}_z^j),
\end{equation}
for $j=1,2$, where $\Delta\theta_{1(2)}$ is the size of one step for
walker 1(2) which depends on the effective coupling $\chi$ and the
time duration $t_{1(2)}$.

\subsection{The Coin Tosses}
\subsubsection{The Separable Coin Toss}

The coin tosses can be implemented on charge qubits with
cavity-assisted interactions. Now we turn off the couplings between
the charge qubits and the horizontal cavities. Then we turn on the
coupling between the charge qubits and the vertical cavity
subsequently. The time-dependent driving field applying on the
vertical cavity~\cite{Bla04}
\begin{equation}
\label{eq:Hd} \hat{H}_\text{d}=\epsilon(t)\left(\hat{b}^\dagger
\text{e}^{-i\omega_\text{d}t}
            +\hat{b}\text{e}^{i\omega_\text{d}t}\right)
\end{equation}
can be used to implement the separable coin toss. It is sufficient
to let $\epsilon(t)$ be a square wave so $\epsilon$ is a constant
($\epsilon=0$ when the field is off). In the dispersive regime and
in a frame rotating at $\omega_\text{d}$ for the qubit and the
resonator, $\hat {H}_j=\hat{H}_\text{JC}^{j}+\hat{H}_\text{d}$ can
be replaced by the effective Hamiltonian
\begin{align}
\label{eq:Heff1}
     \hat{H}_\text{1q}^{j}=&\chi\hat{b}^\dagger\hat{b}\hat{\sigma}_z^j-\frac{\delta_\text{da}}{2}\hat{\sigma}_z^j
+\frac{\Omega_\text{R}}{2}\hat{\sigma}_x^j-\delta_\text{dc}\hat{b}^\dagger\hat{b}+\epsilon(\hat{b}^\dagger
+ \hat{b})
\end{align}
with
\begin{equation}
\delta_\text{da}=\omega_\text{d}-\omega_\text{a},
\delta_\text{dc}=\omega_\text{d}-\omega_\text{c},
\end{equation}
\begin{equation}
\label{eq:Rabi}
    \Omega_\text{R}=\frac{2g\epsilon}{\delta_\text{dc}}
\end{equation}
the Rabi frequency.

The first term in the above expression effects the coin-induced
walker phase shift. The atom transition is an ac-Stark shifted by
$g^2\hat{b}^\dagger\hat{b}\Delta$. To implement
\begin{equation}
H\otimes
H=\Pi_{j=1,2}\exp\left[it_\text{H}\Omega_\text{R}\hat{\sigma}_x^j/2\right]
\end{equation}
on the coin, we choose
\begin{equation}
\omega_\text{d}=\frac{2\bar{n}g^2}{\Delta}-\frac{2g\epsilon}{\Delta}+\omega_\text{a}
\end{equation}
with pulse duration $t_\text{H}=\pi/2\Omega_\text{R}$.

After the coin flipping we shut off both the external field and the
coupling between the charge qubits and the vertical cavity, and turn
on the coupling between the charge qubits and the horizontal
cavities. The free evolution continues for a duration $t_j$ for
charge qubit $j$ for the conditional phase shifts.

\subsubsection{The $\sqrt{i\text{SWAP}}$ Coin Toss}

If one turn on the couplings between the two charge qubits and the
vertical cavity at the same time, the effective Hamiltonian of
$\hat{H}_\text{JC}$ can be written as
\begin{align}
\label{eq:Heff2}
     \hat{H}_\text{2q}=&\big(\omega_\text{c}+\chi\sum_{j=1,2}\hat{\sigma}_z^j\big)\hat{b}^\dagger\hat{b}
     +\frac{1}{2}\big(\omega_\text{a}+\chi\big)\sum_{j=1,2}\hat{\sigma}_z^j\nonumber\\
&+\chi\big(\hat{\sigma}_+^1\hat{\sigma}_-^2+\hat{\sigma}_-^1\hat{\sigma}_+^2\big).
\end{align}
The forth term is the induced dipole-dipole interaction between the
two charge qubits, which can be used to implement the
$\sqrt{-i\text{SWAP}}$ coin toss on the charge states
\begin{equation}
\sqrt{-i\text{SWAP}}= \left(
         \begin{array}{cccc}
           1 & 0 & 0 & 0 \\
           0 & \cos\theta & -i\sin\theta & 0 \\
           0 & -i\sin\theta& \cos\theta & 0 \\
           0 & 0 & 0 & 1 \\
         \end{array}
       \right)
\end{equation}
with $\theta=\chi t_\text{s}$ and $t_\text{s}$ the evolution time of
(\ref{eq:Heff2}) \begin{equation}
U_{2q}=\exp\left[-i\theta\big(\hat{b}^\dagger\hat{b}+\frac{1}{2}\big)\sum_{j=1,2}\hat{\sigma}_z^j\right]\sqrt{i\text{SWAP}}.
\end{equation}



\subsubsection{The DFT Coin Toss}

The similar system can be used to implement a 2D discrete Fourier
transform (DFT) coin toss, defined in Eq.~(\ref{eq:DFT}). Note that
the Hadamard transformation is the 1D DFT~\cite{imp1}.

For the coin flipping, if we choose the duration $t_j$ to increase
geometrically with qubit number as $\chi t_j=2^j\pi/4$, then the
Hamiltonian
$\hat{H}_\text{int}^{j}=\chi\hat{b}^\dagger\hat{b}\hat{\sigma}_z^j$
generate a unitary transformation
\begin{equation}
D=\exp\big(\frac{-i\pi\hat{b}^\dagger\hat{b}\hat{\Upsilon}}{2}\big),
\end{equation}
where the electronic operator $\hat{\Upsilon}$ provides a binary
ordering of the qubits:
\begin{equation}
\hat{\Upsilon}=\sum_{j=1,2}2^{j-1}\hat{\sigma}_z^j.
\end{equation}
The eigenvectors of the operator $\hat{\Upsilon}$ are the electronic
number states
\begin{equation}
\hat{\Upsilon}=\sum_{k=0}^{3}k\ket{k}\bra{k},
\end{equation}
where
$\ket{k}=\ket{S_N}_N\otimes\ket{S_{N-1}}_{N-1}\otimes...\otimes\ket{S_1}_1$,
$S_j=0,1$ and $k=S_N\time 2^{N-1}+S_{N-1}\times
2^{N-2}+...+S_1\times 2^0$. The binary expansion for $k$ is thus
just the string $S_NS_{N-1}...S_1$.

This unitary operation can be used to implement a 2D DFT shown in
Eq.~(\ref{eq:DFT}). First we turn on the coupling between charge
qubit $1$ and the vertical cavity and after time duration $t_1$ it
is turned off. Then we turn on that between charge qubit $2$ and the
vertical cavity and after time duration $t_2$ it is turned off. Thus
the DFT operation on two charge qubits is realized.

\subsubsection{The Grover Coin Toss}

The induced dipole-dipole interaction between two charge qubits can
be used to implement the Grover coin toss~\cite{imp2}. Now we turn
on the coupling between the charge qubits and the vertical cavity at
the same time and apply a drive field (\ref{eq:Hd}) on the vertical
cavity. The effective Hamiltonian of the system in the dispersive
limit becomes
\begin{align}
\label{eq:Heff}
     \hat{H}'_\text{2q}=&\sum_{j=1,2}\big(\chi\hat{b}^\dagger\hat{b}\hat{\sigma}_z^j-\frac{\delta_\text{da}}{2}\hat{\sigma}_z^j
+\frac{\Omega_R}{2}\hat{\sigma}_x^j\big) \nonumber \\
&+\chi\big(\hat{\sigma}_+^1\hat{\sigma}_-^2+\hat{\sigma}_-^1\hat{\sigma}_+^2\big)-\delta_\text{dc}\hat{b}^\dagger\hat{b}+\epsilon(\hat{b}^\dagger
+ \hat{b}).
\end{align}
If $\Omega_\text{R}\gg \delta_\text{da},\delta_\text{dr},g$, we can
get the time evolution of the system in the interaction picture
\begin{equation}
U_\text{I}(t)=e^{-i\hat{H}_0t}e^{-i\hat{H}_et}
\end{equation}
with
\begin{align}
&\hat{H}_0=\Omega_\text{R}/2\sum_{j=1,2}\hat{\sigma}_x^j\\
&\hat{H}_e=\chi\big(\hat{\sigma}^\dagger_1\hat{\sigma}^\dagger_2+\hat{\sigma}^\dagger_1\hat{\sigma}^-_2\big)+h.c..\nonumber
\end{align}
For choosing $\chi t=\pi/8$ and $\Omega_\text{R}/\chi=16m+4$ for $m$
an integer, we can get
\begin{equation}
U_\text{I}(t)=-G.
\end{equation} So by choosing appropriate values of parameters, we
can generate a two-qubit Grover operation on two charge qubits.

The multi-walker QWs can be implemented by three steps as follows.
First, the cavity grid is prepared in the initial state
$\ket{\Psi}_\text{ini}$. Second, we turn off the couplings between
the charge qubits and the corresponding horizontal cavity modes and
turn on those between the charge qubits and vertical cavity. We
applies a coin flipping operation on the charge state. Third, we
turn off the coupling between the charge qubits and vertical cavity
and turn of those with horizontal ones and the free evolution of the
interaction between the charge qubits and cavity modes is used to
implement the conditional phase shift for one step of QW.

\section{open system}

Coupling to additional uncontrollable degree of freedom leads to
energy relaxation and dephasing in the system. In the Born-Markov
approximation, these effects can be characterized by a cavity photon
leakage rate $\kappa$ and a pure dephasing rate $\gamma$ for each
qubit of the coin state. The open system thus evolves according to
\begin{equation}
\frac{\partial\rho}{\partial
t}=-i\big[\hat{H}_\text{int},\rho\big]+\sum_{j=1,2}\kappa_j \mathcal
{D}\big[\hat{a}_j\big]\rho+\frac{\gamma_j}{2}\mathcal{D}\big[\hat{\sigma}^j_z\big]\rho
\end{equation}
with
\begin{equation}
\mathcal{D}\big[\hat{L}\big]\rho\equiv\frac{1}{2}\big(2\hat{L}\rho\hat{L}^\dagger
-\hat{L}^\dagger\hat{L}\rho-\rho\hat{L}^\dagger\hat{L}\big).
\end{equation}

The master equation is used to compute $\rho(t)$ from which the
reduced state of the walker $\rho_\text{w}=\text{Tr}_\text{c}\rho$
is obtained. As a signature of QWs the phase distribution can be
obtained by performing full optical homodyne tomography on the
cavity to obtain the Wigner function, and hence the standard
deviation thereby can be determined.

\begin{figure}[tbp]
   \includegraphics[width=8.25cm]{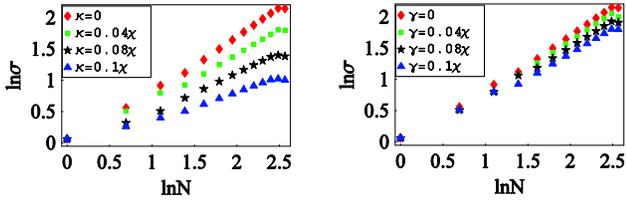}
   \caption{(Color online.) The ln-ln plot of the phase spreads $\sigma(\phi)$ of the
   three-side coin QWs implemented in cavity QED with the DFT coin tosses
   and step size $\Delta\theta=0.8$ as functions of the phase $\phi$ for the initial coin states
   $\left(\ket{1}+i\ket{-1}\right)\otimes \left(\ket{1}+i\ket{-1}\right)/2$ with the presence of decoherence.
   (a) Fixing the dephasing rate $\gamma=0.06\chi$, one can observe the QW-RW
   transition by increasing the cavity decay rate from $\kappa=0$ to $\kappa=0.1\chi$.
   (b) Fixing the decay rate $\kappa=0.01\chi$, the slope of ln-ln plot decreases slowly
   with $\gamma/\chi$ increasing.}
   \label{figure3}
\end{figure}

\begin{figure}[tbp]
   \includegraphics[width=8.25cm]{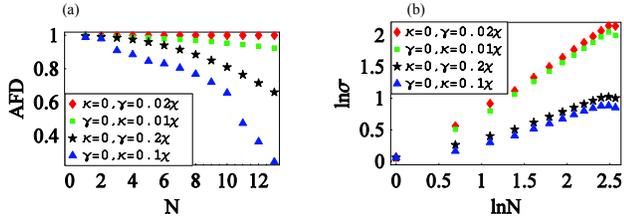}
   \caption{(Color online.) With the step size $\Delta\theta=0.8$, (a) AFDs and
   (b) ln-ln plot of the phase spreads $\sigma(\phi)$ of the
   three-side coin QWs implemented in cavity QED with the DFT coin tosses
   as functions of the phase $\phi$ for the initial coin states
   $\left(\ket{1}+i\ket{-1}\right)\otimes \left(\ket{1}+i\ket{-1}\right)/2$ in the presence of decoherence.}
   \label{figure4}
\end{figure}

With the realistic system parameters~\cite{Bla04,Bla07}
$\left(\omega_\text{a},\omega_\text{c},g,\epsilon\right)/2\pi=\left(7000,5000,100,1000\right)$MHz,
simulated evolutions of the standard deviations of the phase
distribution for the first several steps are presented in Fig.~2,
which clearly reveal slope compatible with the characteristic
quadratic decrease in phase spreading for increasing decoherence of
the two-walker QW until the transition to the RW \cite{note}. Here
we use a DFT coin QW with initial coin state
$\left(\ket{1}+i\ket{-1}\right)\otimes\left(\ket{1}+i\ket{-1}\right)/2$
as an example with the decay rate of cavity $\kappa_1=\kappa_2$
increasing from $0$ to $0.1\chi$ and the dephasing rate
$\gamma_1=\gamma_2=0.06\chi$ fixed. Thus the QW-RW transition is
observed in Fig.~2(a). Furthermore, if the decay rate
$\kappa_1=\kappa_2=0.01\chi$ is fixed, with the dephasing rate
increasing the standard deviation of the phase distribution $\sigma$
as a function of the number of steps $N$ decreases slowly. And with
$\gamma$ increasing from $0$ to $0.1\chi$ the slope of the
$\ln\sigma$-$\ln N$ plot decreases from $0.970$ to $0.810$ in
Fig.~2(b).

Except for the standard deviation of the probability distribution,
one can determine the effects of the decoherence on the quantum
behavior of QWs via the AFD. As shown in Fig.~3(a), we have
numerically calculated the AFD of the DFT coin QW with the initial
coin state
$\left(\ket{1}+i\ket{-1}\right)\otimes\left(\ket{1}+i\ket{-1}\right)/2$
in the two cases: (i) only with the decay of the cavity; (ii) only
with the dephasing of the charge qubits. Using the parameters with
which the standard deviations in the presence of the two sources of
decoherence are quite close shown in Fig.~3(b), one can obverse that
the AFDs of the two cases are quite different and the difference
increases with the decoherence increasing. In Fig.~3(b), the
$\ln\sigma$-$\ln N$ plots are shown with different parameters. The
result with $\kappa_1=\kappa_2=0$ and $\gamma_1=\gamma_2=0.02\chi$
($\kappa_1=\kappa_2=0$ and $\gamma_1=\gamma_2=0.2\chi$) is quite
similar with that with $\gamma_1=\gamma_2=0$ and
$\kappa_1=\kappa_2=0.01\chi$ ($\gamma_1=\gamma_2=0$ and
$\kappa_1=\kappa_2=0.1\chi$). Whereas, in Fig.~3(a) with the same
choices of parameters, the AFDs are quite different and derive from
each other. The AFD is more sensitive to the cavity decay rather
than to the charge dephasing. These results imply the significant
effect on the QW-RW transition from the cavity decay $\kappa$.
Moreover, the cavity decay $\kappa$ is much more important than the
charge dephasing $\gamma$ with respect to the scaling of AFD with
time $t$ (proportional to the number of steps $N$). The dephasing
rate $\gamma$ mainly leads to smearing the phase distribution and
the phase distribution loses its symmetry.

\section{conclusion}

In summary, we have introduced a protocol to implement a QW with a
multi-walker in phase space using cavity QED. The implementation of
a QW with a multi-walker opens up the interesting possibility to
introduce entanglement and more advanced walks. With different coin
tosses and initial states the multi-walker QWs show different
probability distributions which deviate strongly from the RWs and
show faster spreadings and localization effect. By introducing
decoherence, the transmission from QWs to the classical versions is
observed. We propose a physical realization to investigate the
decoherence-induced irreversibility of QWs via the AFD. Our scheme
provides an experimentally available way to monitor the detrimental
influence on the QW by different decoherence sources. It is observed
that the cavity decay has more detrimental effects on QWs rather
than the dephasing of the charge qubits, which allows us to
understand better the QW simulation in a realistic system. In
conclusion, our theory establishes a pathway to realizing a
many-step QW with multi-walker, and our techniques for observing the
signature of QW via the AFD would be useful for general quantum
information protocols.

\begin{acknowledgments}
This work has been supported by the National Natural Science
Foundation of China under Grant Nos 11004029 and 11174052, the
Natural Science Foundation of Jiangsu Province under Grant No
BK2010422, the Ph.D. Programs Foundation of Ministry of Education of
China, the Excellent Young Teachers Program of Southeast University
and the Major State Basic Research Development Program of China (973
Program) under Grant No 2011CB921203.
\end{acknowledgments}


\begin{references}

\bibitem{QW} Y. Aharonov, L. Daviovich and N. Zagury, \pra
\textbf{48}, 1687 (1993).

\bibitem{algorithm} D. Aharonov, A. Ambainis, J. Kempe and U.
Vazirani, Proc. 33th STOC (New York) pp 50-59 (2000).

\bibitem{circuit} M. A. Nielsen and I. L. Chuang, Quantum Computation
and Quantum Information (Cambridge: Cambridge University Press)
(2000).

\bibitem{oneway} R. Raussendorf and H. J. Briegel, \prl \textbf{86},
5188 (2001).

\bibitem{con}A. M. Childs, \prl \textbf{102}, 180501 (2009).

\bibitem{dis}N. B. Lovett, S. Cooper, M. Everitt, M. Trevers and V.
Kendon, \pra \textbf{81}, 042330 (2010).

\bibitem{qo} P. Zhang, B. H. Liu, R. F. Liu, H. R. Li, F. L. Li and G. C. Guo, \pra \textbf{81}, 052322 (2010).

\bibitem{ion1} B. C. Travaglione and G. J. Milburn, \pra \textbf{65},
032310 (2002).

\bibitem{ion2} P. Xue, B. C. Sanders and D. Leibfriend, \prl \textbf{103}, 183602 (2009).

\bibitem{cavity1} P. Xue, B. C. Sanders, A. Blais and K. Lalumi\'{e}re, \pra
\textbf{78}, 042334 (2008).

\bibitem{cavity2}P. Xue and B. C.
Sanders, New J. Phys. \textbf{10}, 053025 (2008).

\bibitem{ol} W. D\"{u}r, R. Raussendorf, V. Kendon and H. J.
Briegel, \pra \textbf{66}, 052319 (2002).

\bibitem{expsp}M. A. Broome, A. Fedrizzi, B. P. Lanyon, I. Kassal,
A. Aspuru-Guzik and A. G. White, \prl \textbf{104}, 153602 (2010).

\bibitem{expion}F. Zahringer, G. Kirchmair, R. Gerritsma, E. Solano,
R. Blatt and C. F. Roos, Phys. Rev. Lett. \textbf{104}, 100503
(2010).

\bibitem{expatom} M. Karski, L. F\"{o}rster, J.-M. Choi, A.
Steffen, W. Alt, D. Meschede and A. Widera, Science \textbf{325},
174 (2009).

\bibitem{Mackay02} T. D. Mackay, S. D. Bartlett, L. T. Stephenson and B. C.
Sanders, J. Phys. A: Math. Gen. \textbf{35}, 2745 (2002).

\bibitem{more1} A. Peruzzo, M. Lobino, J. C. F. Matthews, N. Matsuda, A. Polliti,
K. Poulios, X. Q. Zhou, Y. Lahini, N. Ismail, K. W\"{o}rhoff, Y.
Bromberg,Y. Silberberg, M. G. Thompson and J. L. O'Brien, Science
\textbf{329}, 1500 (2010).

\bibitem{more2} T. A. Brun, H. A. Carteret and A. Ambainis, \pra \textbf{67},
052317 (2003).

\bibitem{more3} Y. Omar, N. Paunkovic, L. Sheridan and S. Bose, Phys. Rev. A
\textbf{74}, 042304 (2006).

\bibitem{more4} C. Liu and N. Petulante,
\pra \textbf{79}, 032312 (2009).

\bibitem{more5} P. P. Rohde, A. Schreiber, M. Stefanak, I. Jex and C. Silberhorn, New J. Phys. {\bf13},
013001 (2011).

\bibitem{more6} M. Stefanak, S. M. Barnett, B. Kollar, T. Kiss and I. Jex, New J. Phys.
{\bf13}, 033029 (2011).

\bibitem{more7} S. D. Berry and J. B. Wang, \pra {\bf 83}, 042317 (2011).

\bibitem{more8} P. Xue and B. C. Sanders, \pra {\bf 85} 022307 (2012).

\bibitem{Feng} Y. Y. Xu, F. Zhou, L. Chen, Y. Xie, P. Xue and M.
Feng, unpublished.

\bibitem{AFD} F. M. Cucchietti, D. A. R. Dalvit, J. P. Paz and W. H.
Zurek, Phys. Rev. Lett. \textbf{91}, 210403 (2003).

\bibitem{Holevo} A. S. Holevo, Lect. Notes Math. {\bf1055}, 153 (1984).

\bibitem{Keating} J. P. Keating, N. Linden, J. C. F. Matthews and A. Winter, Phys. Rev. A {\bf76}, 012315
(2007).

\bibitem{SC} A. Schreiber, K. N. Cassemiro, V. Poto\v{c}ek,
A. G\'{a}bris, I. Jex and C. Silberhorn, \prl {\bf 106}, 180403
(2011).

\bibitem{grrid}F. Helmer, M. Mariantoni, A. G. Fowler, J. von Delft, E. Solano and F. Marquardt, Europhysics Letters \textbf{85} 50007
(2009).

\bibitem{Bla04}
A. Blais, R. S. Huang, A. Wallraff, S. M. Girvin and R. J.
Schoelkopf, \pra \textbf{69}, 062320 (2004).

\bibitem{Bla07} A. Blais, J. Gambetta, A. Wallraff, D. I. Schuster, S. M.
Girvin, M. H. Devoret and R. J. Schoelkopf, \pra \textbf{75}, 032329
(2007).

\bibitem{Xue} P. Xue, Phys. Lett. A {\bf 374}, 2601 (2010); P.
Xue, Chin. Phys. Lett. {\bf 28}, 070305 (2011); P. Xue, Chin. Phys.
B {\bf 20}, 100310 (2011).

\bibitem{imp1} H. F. Wang, A. D. Zhu, S. Zhang and K. H. Yeon, New J. Phys. {\bf 13}, 013021 (2011).

\bibitem{imp2} W. L. Yang, C. Y. Chen and M. Feng, Phys. Rev. A {\bf 76},
054301 (2007).

\bibitem{note} In this Sec., we only consider short time limit. That is
we focus on the behavior of QW with decoherence for the first few
steps.

\end{references}
\end{document}